\pdfoutput=1
\documentclass[preprint,showpacs,preprintnumbers,amsmath,amssymb,floatfix,endfloats*]{revtex4}

\usepackage{graphicx}
\usepackage{dcolumn}
\usepackage{bm}
\usepackage{amsfonts}

\DeclareMathAlphabet{\mathsfsl}{OT1}{cmr}{bx}{it}
\begin{document}
\title{Collective nonaffine displacements in amorphous materials
during large-amplitude oscillatory shear}
\author{Nikolai V. Priezjev}
\affiliation{Department of Mechanical and Materials Engineering,
Wright State University, Dayton, OH 45435}
\date{\today}
\begin{abstract}

Using molecular dynamics simulations, we study the transient
response of a binary Lennard-Jones glass subjected to periodic shear
deformation.   The amorphous solid is modelled as the
three-dimensional Kob-Andersen binary mixture at a low temperature.
The cyclic loading is applied to slowly annealed, quiescent samples,
which induces irreversible particle rearrangements at large strain
amplitudes, leading to stress-strain hysteresis and a drift of the
potential energy towards higher values.  We find that the initial
response to cyclic shear near the critical strain amplitude involves
disconnected clusters of atoms with large nonaffine displacements.
In contrast, the amplitude of shear stress oscillations decreases
after a certain number of cycles, which is accompanied by the
initiation and subsequent growth of a shear band.

\end{abstract}




\maketitle

\section{Introduction}

The development of advanced thermomechanical processing techniques
for metallic glasses might provide access to a broader range of
states and thus permit exploitation of the improved material
properties~\cite{Greer16}. It has been long realized that an
elementary plastic event in deformed glasses involves a collective
rearrangement of a small group of atoms, known as a shear
transformation zone~\cite{Spaepen77,Argon79,Falk98}. Recently, it
was argued that the mechanical yield in amorphous solids does not
necessarily result in structural changes but rather represents a
transition from a constrained set to a vast number of available
configurations~\cite{Procaccia16}. At sufficiently small shear
rates, glassy materials exhibit shear localization in the form of
fluidized shear bands running across the
sample~\cite{VarnikBB03,Falk06}.  Under a constant strain rate, the
plastic flow follows the formation of a percolating cluster of
mobile regions characterized by large nonaffine
displacements~\cite{Horbach16,HorbachJR16}. However, shear band
initiation and evolution during more complex time-dependent
deformation protocols remain relatively unexplored.

\vskip 0.05in

In the last several years, the microscopic mechanism of deformation
of amorphous materials during periodic shear was studied extensively
using atomistic
simulations~\cite{Priezjev13,Sastry13,Reichhardt13,HernHoy13,Priezjev14,IdoNature15,Priezjev16,Sastry16,Priezjev16a,Kawasaki16,OHern16}
and experimental
measurements~\cite{Chaikin08,BiroliPRL09,Petekidis13,Arratia13,Cipelletti14,Bartolo14,Echoes14,Ganapathy14,Zaccone16,Manneville16}.
In particular, it was shown that during cyclic loading at small
strain amplitudes below the yielding transition, particle
trajectories remain reversible over consecutive
cycles~\cite{Sastry13,Reichhardt13,HernHoy13,IdoNature15}, and,
notably, some atoms undergo repetitive nonaffine displacements with
amplitudes that are comparable to the cage
size~\cite{Zaccone16,Priezjev16,Priezjev16a}.   It was also found
that more slowly cooled glasses undergo smaller particle
rearrangements and become more reversible under athermal,
quasistatic cyclic shear~\cite{OHern16}.   In the absence of thermal
fluctuations, a sharp onset of particle diffusion is detected at the
critical strain
amplitude~\cite{Sastry13,Reichhardt13,HernHoy13,IdoNature15,Sastry16},
which can be interpreted as a nonequilibrium first-order dynamic
phase transition~\cite{Kawasaki16}.    Rather interestingly, the
results of numerical simulations indicate that the yielding
transition in oscillatory shear is not accompanied by changes in
microscopic structure~\cite{Kawasaki16}.

\vskip 0.05in

At finite temperatures, nonaffine rearrangements in glasses arise
both from thermal motion of atoms and from the imposed shear
deformation, and, thus, shear-induced activation of irreversible
structural rearrangements is assisted by thermal
noise~\cite{Horbach16,Priezjev16a}.   Under cyclic loading, atoms
with large nonaffine displacements are spatially organized into
clusters, which become comparable with the system size near the
yield strain~\cite{Priezjev16,Priezjev16a}.     Moreover, upon
increasing strain amplitude, spatial correlations of nonaffinity
become extended over longer range, although they remain present even
in the absence of mechanical deformation due to thermal
fluctuations~\cite{Priezjev16a}.    It was further shown that over
many deformation cycles, the structural relaxation dynamics is
spatially and temporally heterogeneous; namely, mobile particles, or
cage jumps, tend to aggregate into transient
clusters~\cite{Priezjev13,Priezjev14}.    Furthermore, during cyclic
shear near the yield strain, a cluster of atoms with large
reversible nonaffine displacements induces a long-range,
time-dependent elastic field that in turn might trigger secondary
structural rearrangements. This situation was considered separately
in recent studies~\cite{Priezjev15,Priezjev15a}, where it was found
that a local reversible shear transformation in a quiescent system
induces irreversible cage jumps, and their density is larger in the
cases of weaker damping or slower shear transformation.

\vskip 0.05in

More recently, the fatigue mechanism in bulk metallic glasses
subjected to tension-compression cyclic loading was investigated
using molecular dynamics (MD) simulations~\cite{GaoNano15} and
finite element modeling~\cite{JiangFEM16}.   It was found that the
amplitude of stress oscillations is reduced after the first several
cycles, which is associated with the formation of a dominant shear
band across the whole system~\cite{GaoNano15,JiangFEM16}.   In this
process, the initiation of a shear band is preceded by an
accumulation of shear transformation zones at the surface of the
material.    Moreover, it was shown that higher cycling frequency
leads to a larger number of cycles to
failure~\cite{GaoNano15,JiangFEM16}.   Nevertheless, the microscopic
details of the relaxation process in amorphous materials under
various types of loading conditions remain not fully understood.

\vskip 0.05in

In this paper, molecular dynamics simulations are carried out to
investigate the transient response of a binary glass to
large-amplitude oscillatory shear deformation.  It will be shown
that above the yield strain, the structural relaxation process
involves irreversible rearrangements of particles within a shear
band, which leads to stress-strain hysteresis and increase of the
potential energy over consecutive cycles.  Near the critical strain
amplitude, the formation of a shear band is delayed for a number of
cycles and it coincides with the increase of the potential energy
and a distinct drop in the shear stress amplitude.

\vskip 0.05in

The rest of the paper is organized as follows. The details of
molecular dynamics simulations and the deformation protocol are
described in the next section. The variation of the potential energy
and shear stress, as well as stress-strain hysteresis and spatial
configurations of particle with large nonaffine displacements at
different strain amplitudes are presented in
Sec.\,\ref{sec:Results}.  The results are briefly summarized in the
last section.

\section{Molecular dynamics simulations}
\label{sec:MD_Model}

\begin{figure}[t]
\includegraphics[width=10.cm,angle=0]{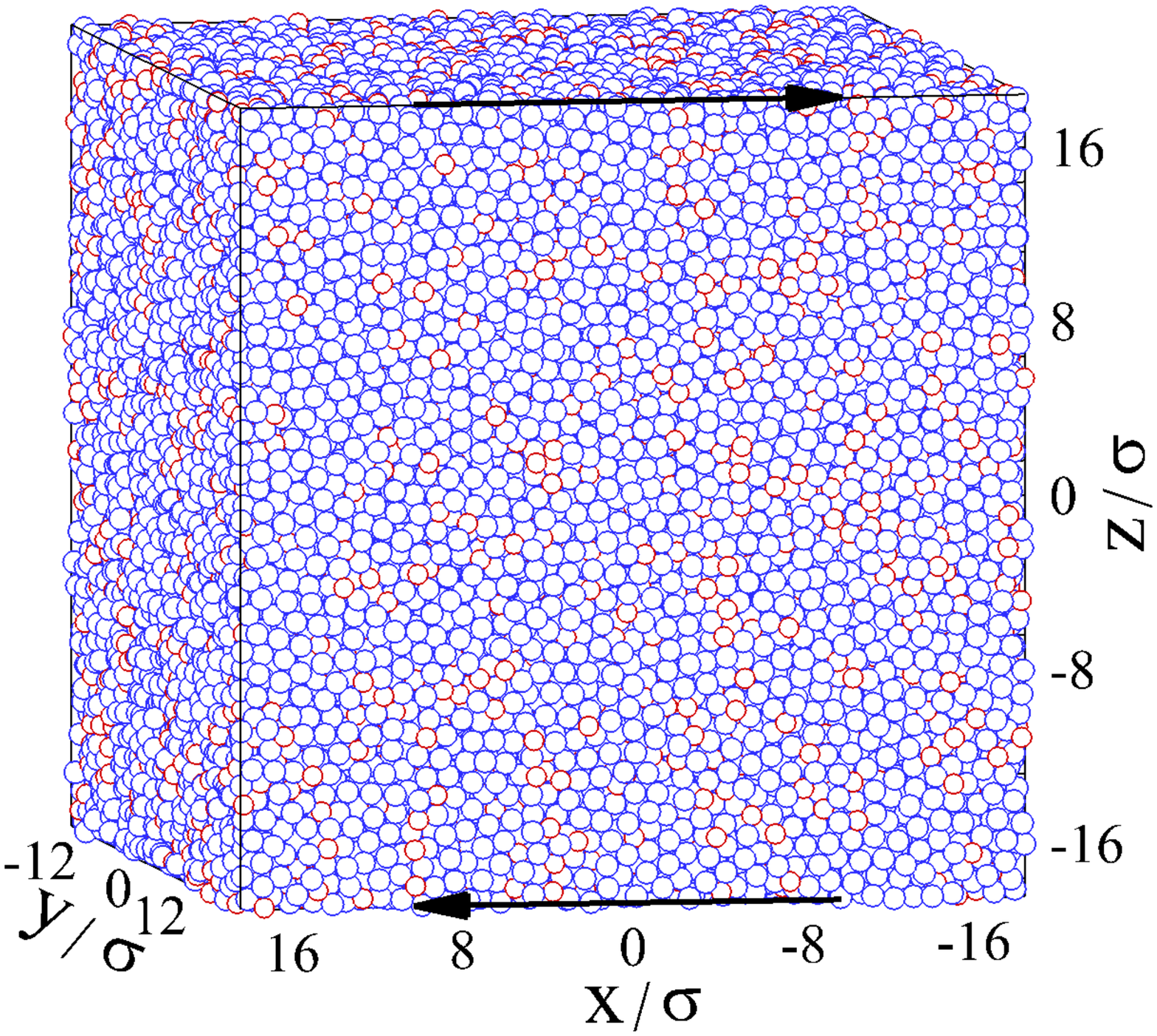}
\caption{(Color online) A snapshot of the instantaneous atomic
configuration of the binary Lennard-Jones glass at the temperature
$T_{LJ}=10^{-2}\,\varepsilon/k_B$ during periodic shear deformation
along the $xz$ plane (indicated by the black arrows) with the strain
amplitude $\gamma_0=0.10$.  The atoms of type A are denoted by large
blue circles and atoms of type B are shown by small red circles.
Atoms are not drawn to scale. The total number of atoms is
$N=60\,000$. }
\label{fig:snapshot_system}
\end{figure}

In order to study the deformation dynamics of amorphous materials,
we consider the standard Kob-Andersen (KA) (80:20) binary mixture
model~\cite{KobAnd95} that was originally designed to reproduce the
properties of the metal alloy
$\text{Ni}_{80}\text{P}_{20}$~\cite{Weber85}.   In our setup, the
system consists of $N_A=48\,000$ large atoms of type A and
$N_B=12\,000$ small atoms of type B that are confined in a
three-dimensional periodic cell.  A snapshot of the equilibrated
system of $N=60\,000$ atoms is presented in
Fig.\,\ref{fig:snapshot_system}.   In the KA model, any two atoms
$\alpha,\beta=A,B$ interact via the truncated Lennard-Jones (LJ)
potential as follows:
\begin{equation}
V_{\alpha\beta}(r)=4\,\varepsilon_{\alpha\beta}\,\Big[\Big(\frac{\sigma_{\alpha\beta}}{r}\Big)^{12}\!-
\Big(\frac{\sigma_{\alpha\beta}}{r}\Big)^{6}\,\Big],
\label{Eq:LJ_KA}
\end{equation}
where the interaction parameters are fixed to
$\varepsilon_{AA}=1.0$, $\varepsilon_{AB}=1.5$,
$\varepsilon_{BB}=0.5$, $\sigma_{AB}=0.8$, $\sigma_{BB}=0.88$, and
$m_{A}=m_{B}$~\cite{KobAnd95}.  This choice of parameters defines a
highly non-additive LJ potential that prevents crystallization at
low temperatures~\cite{KobAnd95}.   The cutoff radius is
$r_{c,\,\alpha\beta}=2.5\,\sigma_{\alpha\beta}$ and the units of
length, mass, energy, and time are set to $\sigma=\sigma_{AA}$,
$m=m_{A}$, $\varepsilon=\varepsilon_{AA}$, and
$\tau=\sigma\sqrt{m/\varepsilon}$, respectively.  The equations of
motion for each atom were integrated using the Verlet
algorithm~\cite{Allen87,Lammps} with the time step $\triangle
t_{MD}=0.005\,\tau$.  The molecular dynamics simulations were
conducted using the efficient parallel program LAMMPS developed at
Sandia National Laboratories~\cite{Lammps}.

\vskip 0.05in


Each individual sample was carefully prepared by first placing the
atoms in the cubic box and assigning random velocities at the
temperature $1.1\,\varepsilon/k_B$, which is above the computer
glass transition temperature
$T_g\approx0.45\,\varepsilon/k_B$~\cite{KobAnd95}. Here $k_B$ is the
Boltzmann constant. The dimensions of the cubic box of linear size
$L=36.84\,\sigma$ were kept constant in all simulations, and the
corresponding density is $\rho=\rho_{A}+\rho_{B}=1.2\,\sigma^{-3}$.
Second, in the absence of mechanical deformation, the temperature of
the system was gradually reduced with a computationally slow rate of
$10^{-5}\,\varepsilon/k_B\tau$ to the final temperature
$T_{LJ}=10^{-2}\,\varepsilon/k_B$. The data were collected in five
independent samples.

\vskip 0.05in


Followed by the equilibration procedure, the material was subjected
to periodic shear deformation along the $xz$ plane as shown in
Fig.\,\ref{fig:snapshot_system}.   The applied shear strain was
varied periodically according to
\begin{equation}
\gamma(t)=\gamma_{0}\,\,\textrm{sin}(2\pi t / T),
\label{Eq:strain}
\end{equation}
where $\gamma_{0}$ is the strain amplitude and $T$ is the
oscillation period. In what follows, the oscillation period was
fixed to $T=5000\,\tau$ and, correspondingly, the oscillation
frequency is $\omega=2\pi/T=1.26\times 10^{-3}\,\tau^{-1}$.  The
Lees-Edwards periodic boundary conditions~\cite{Allen87} were
employed in the $xz$ plane.   The non-equilibrium MD simulations
were performed in the constant NVT ensemble, where the temperature
was controlled by the dissipative particle dynamics (DPD)
thermostat~\cite{Soddemann03}.   The DPD thermostat is based on the
relative atom velocities and thus the particle dynamics is not
coupled to a flow profile, and, as a result, the formation of shear
bands or other flow inhomogeneities are not suppressed during
deformation of the material~\cite{Soddemann03,Horbach16}.

\section{Results}
\label{sec:Results}

\begin{figure}[t]
\includegraphics[width=12.cm,angle=0]{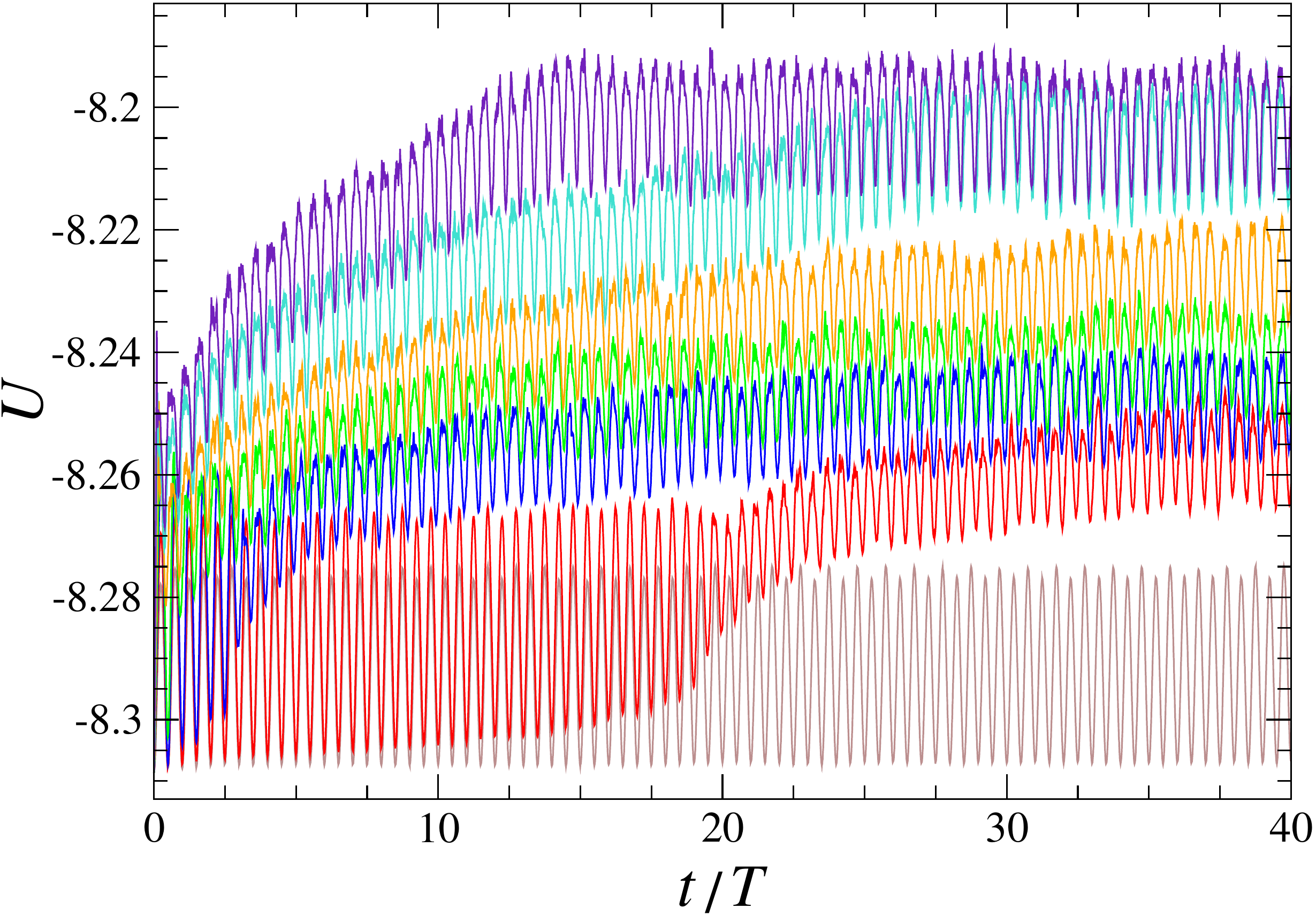}
\caption{(Color online) The variation of the potential energy per
particle $U$ (in units of $\varepsilon$) in one sample during the
first 40 oscillation cycles for the strain amplitudes $\gamma_{0} =
0.07$, $0.08$, $0.09$, $0.10$, $0.12$, $0.16$, and $0.20$ (from
bottom to top). The oscillation period is $T=5000\,\tau$. }
\label{fig:poten}
\end{figure}

The initial response of amorphous materials to mechanical
deformation depends strongly on the preparation
history~\cite{Greer16}.   In particular, it is well recognized that
more slowly annealed glasses acquire a state with a lower potential
energy and exhibit higher yield stress~\cite{Greer16}.    On the
other hand, it was shown that large strain cycles cause irreversible
relaxations in the material and relocate the system to shallower
energy minima, thus, leading to rejuvenation~\cite{Lacks04}.    In
the present study, the potential energy per particle in one
representative sample is plotted in Fig.\,\ref{fig:poten} during 40
cycles for the strain amplitudes $\gamma_{0} = 0.07$, $0.08$,
$0.09$, $0.10$, $0.12$, $0.16$, and $0.20$.   It can be seen that at
the strain amplitude $\gamma_{0} = 0.07$, the variation of the
potential energy is periodic with superimposed noise due to thermal
fluctuations. Remarkably, at $\gamma_{0} = 0.08$, the potential
energy first gradually increases on average during about 20 cycles
and then levels off rapidly to a new regime of oscillations with a
smaller amplitude.    This dynamic transition marks the onset of
large-scale irreversible structural rearrangements that will be
identified more directly below.    As shown in
Fig.\,\ref{fig:poten}, with further increasing strain amplitude, the
potential energy per particle increases more rapidly over
consecutive cycles.

\vskip 0.05in

\begin{figure}[t]
\includegraphics[width=12.cm,angle=0]{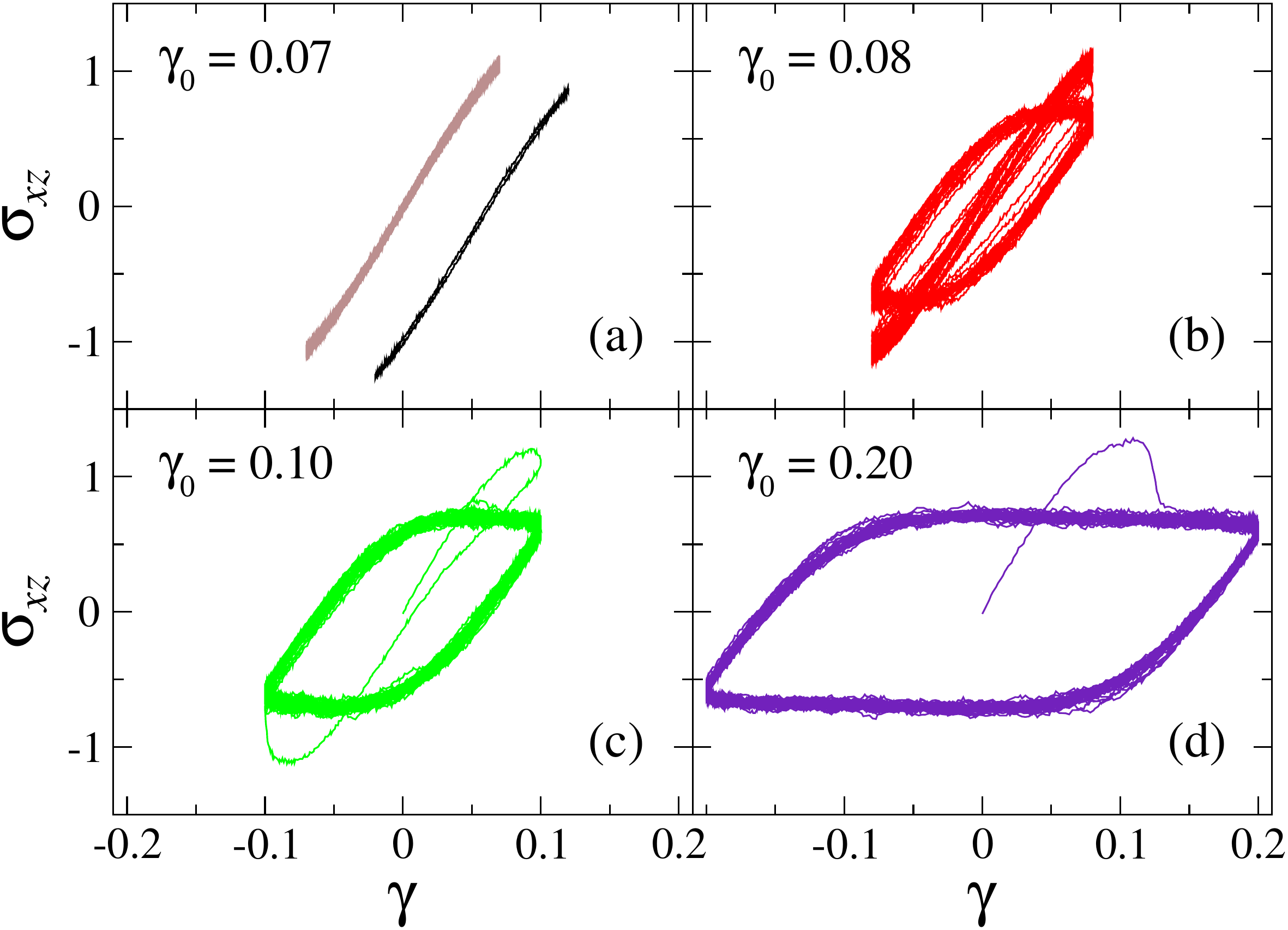}
\caption{(Color online) The shear stress $\sigma_{xz}$ (in units of
$\varepsilon\sigma^{-3}$) versus shear strain $\gamma$ during 40
cycles for the strain amplitudes (a) $\gamma_{0} = 0.07$, (b)
$\gamma_{0} = 0.08$, (c) $\gamma_{0} = 0.10$, and (d) $\gamma_{0} =
0.20$.  The data for the last cycle at $\gamma_{0} = 0.07$ are
included in the panel (a) and displaced for clarity (black curve). }
\label{fig:hysteresis}
\end{figure}

The typical Lissajous curves of shear stress versus strain are
displayed in Fig.\,\ref{fig:hysteresis} for the selected strain
amplitudes $\gamma_{0} = 0.07$, $0.08$, $0.10$, and $0.20$. It can
be seen that a small hysteresis, and, therefore, energy dissipation
appear already at $\gamma_{0} = 0.07$, when the system dynamics is
nearly reversible as suggested by the periodic variation of the
potential energy shown in Fig.\,\ref{fig:poten}. The finite loop
area at $\gamma_{0} = 0.07$ is masked by the initial response and
thermal fluctuations over many cycles. To resolve the hysteresis
curve more clearly, the variation of shear stress as a function of
strain during the last cycle is also presented in
Fig.\,\ref{fig:hysteresis}\,(a). With increasing strain amplitude,
the area of the hysteresis loop increases while the amplitude of
shear stress remains the same (about $0.72\,\varepsilon\sigma^{-3}$)
in steady state.   Most interestingly, in
Fig.\,\ref{fig:hysteresis}\,(b) one can observe a transition from
initially small to large hysteresis after several cycles at the
strain amplitude $\gamma_{0} = 0.08$, which points at enhanced
energy dissipation after about 20 cycles.   Finally, pronounced
shear stress overshoots arise during the first cycle followed by
hysteresis loops with large areas at higher strain amplitudes
$\gamma_{0} = 0.10$ and $0.20$ as shown in
Fig.\,\ref{fig:hysteresis}\,(c) and (d), respectively.

\vskip 0.05in

\begin{figure}[t]
\includegraphics[width=12.cm,angle=0]{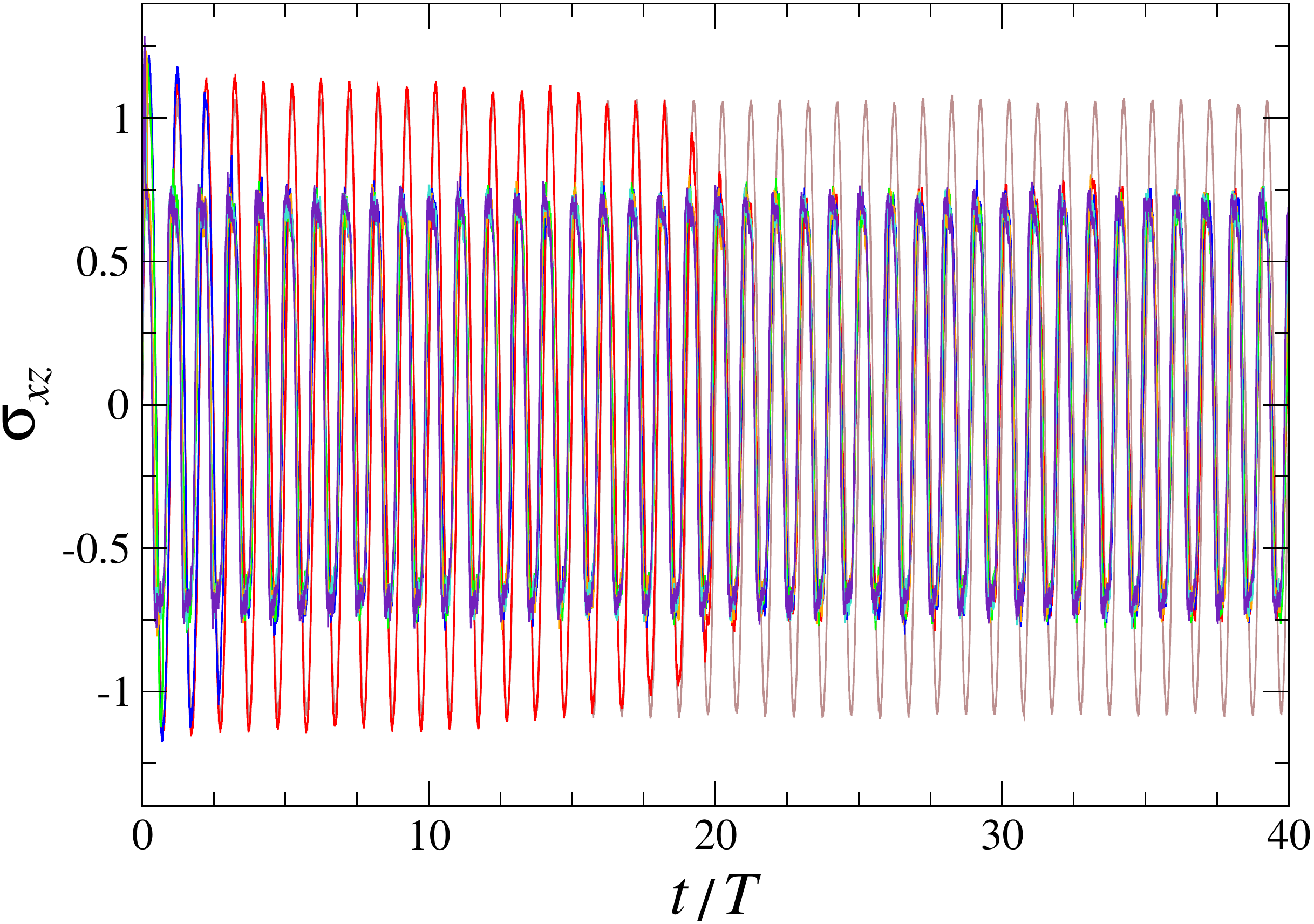}
\caption{(Color online) The shear stress $\sigma_{xz}$ (in units of
$\varepsilon\sigma^{-3}$) during the first 40 oscillation cycles for
the strain amplitudes $\gamma_{0} = 0.07$ \text{(brown)}, $0.08$
\text{(red)}, $0.09$ \text{(blue)}, $0.10$ \text{(green)}, $0.12$
\text{(orange)}, $0.16$ \text{(turquoise)}, and $0.20$
\text{(indigo)}.  The period of oscillation is $T=5000\,\tau$. }
\label{fig:stress_strain_40T}
\end{figure}

The variation of shear stress during forty cycles are presented in
Fig.\,\ref{fig:stress_strain_40T} for different strain amplitudes.
It can be observed that at the strain amplitude $\gamma_{0} = 0.07$,
the amplitude of shear stress oscillations remains unchanged, which
is consistent with the periodic behavior of the potential energy
reported in Fig.\,\ref{fig:poten}.   By contrast, at $\gamma_{0} =
0.08$ in Fig.\,\ref{fig:stress_strain_40T}, the shear stress
amplitude is nearly constant for about 18 cycles and then it
decreases to a smaller value, which occurs at the same time when the
potential energy grows rapidly (see Fig.\,\ref{fig:poten}). This
response to applied periodic shear implies a formation of extended
fluidized regions that determine the stress amplitude of
$0.72\pm0.06\,\varepsilon\sigma^{-3}$.  This value correlates well
with the post-yield shear stress
$\sigma_{SS}=0.68\pm0.04\,\varepsilon\sigma^{-3}$ averaged over five
samples under steady shear at the strain rate
$\dot{\gamma}\tau=10^{-4}$ (not shown).    Furthermore, it can be
seen in Fig.\,\ref{fig:stress_strain_40T} that the shear stress
reaches steady-state oscillations after three cycles at
$\gamma_{0}=0.09$ and after the first cycle at $\gamma_{0} > 0.09$.


\vskip 0.05in


It should be noted that the trends identified in
Figures\,\,\ref{fig:poten}, \ref{fig:hysteresis} and
\ref{fig:stress_strain_40T} were observed in all five independent
samples, although the number of cycles until the dynamic transition
at the strain amplitude $\gamma_{0} = 0.08$ varies from 15 to 59.
These results agree qualitatively with recent numerical simulations
of metallic glasses under tension-compression cyclic loading, where
it was shown that after a certain number of cycles, the stress
amplitude is reduced concomitantly with the initiation of a shear
band across the sample~\cite{GaoNano15,JiangFEM16}.    We also
comment that the transition from reversible dynamics at $\gamma_{0}
= 0.07$ to the plastic regime at $\gamma_{0} = 0.08$ occurs at
higher strain amplitudes than the critical value $\gamma_{0} = 0.06$
reported in the previous MD study~\cite{Priezjev13}, where
simulations were performed at the higher oscillation frequency
$\omega\tau=0.02$ and higher temperature
$T_{LJ}=0.1\,\varepsilon/k_B$.     At the same time, our results are
in agreement with the critical strain amplitude $\gamma_{0} = 0.07$,
which marks the onset of energy dissipation and particle diffusion
in a binary glass under oscillatory athermal quasistatic shear
deformation~\cite{Sastry13,Sastry16}.

\vskip 0.05in

\begin{figure}[t]
\includegraphics[width=12.cm,angle=0]{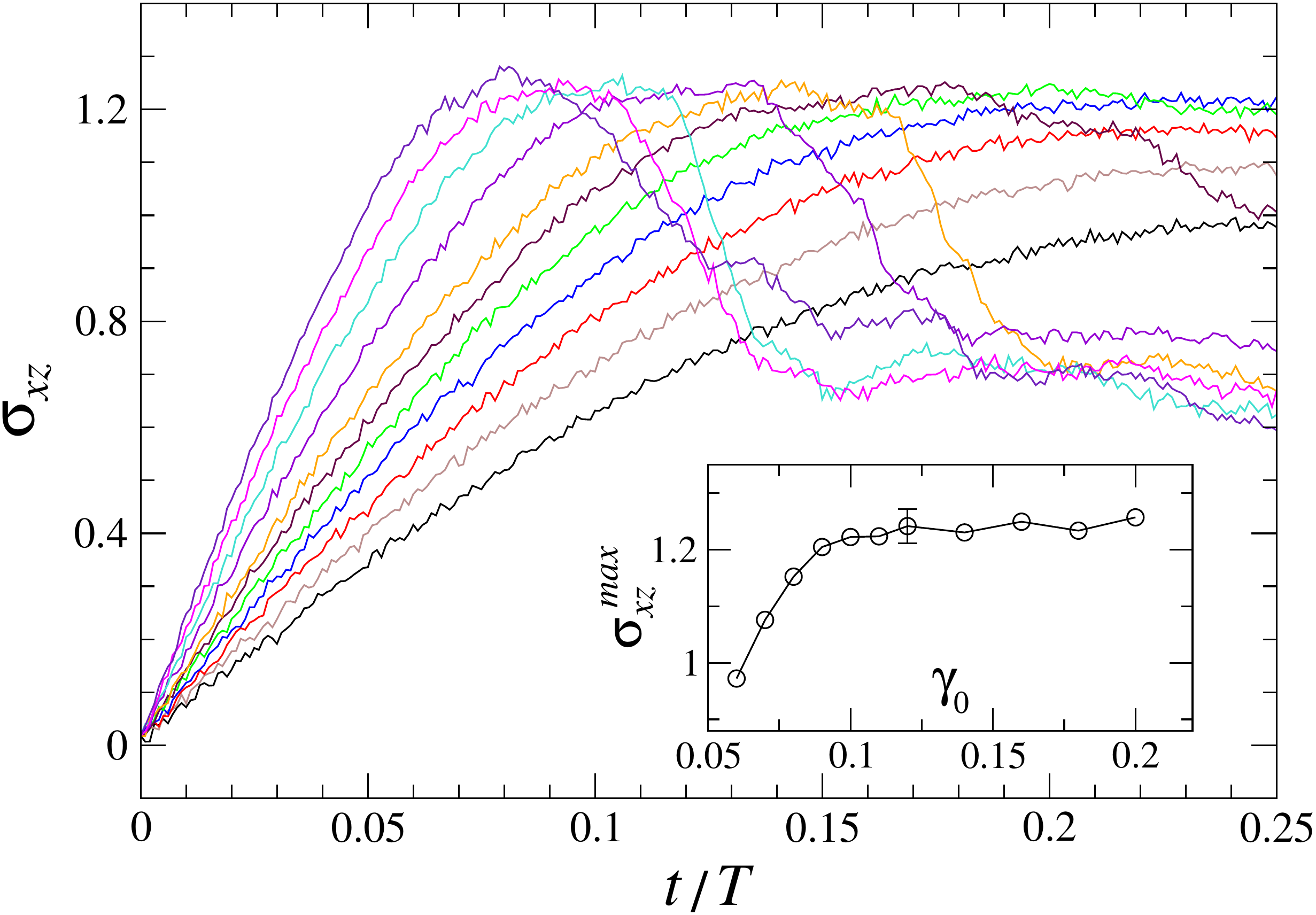}
\caption{(Color online) The shear stress $\sigma_{xz}$ (in units of
$\varepsilon\sigma^{-3}$) during the first cycle for the strain
amplitudes $\gamma_{0} = 0.06$, $0.07$, $0.08$, $0.09$, $0.10$,
$0.11$, $0.12$, $0.14$, $0.16$, $0.18$, and $0.20$ (from right to
left).  The period of oscillation is $T=5000\,\tau$. The data are
taken in one sample.  The inset shows the value of the overshoot
stress averaged over five samples as a function of the strain
amplitude.  }
\label{fig:stress_strain_T4}
\end{figure}

To show the initial response more clearly, the time dependence of
the shear stress during the first quarter of the cycle is presented
in Fig.\,\ref{fig:stress_strain_T4} for different strain amplitudes.
It can be seen that the yield stress becomes apparent at $\gamma_{0}
\geqslant 0.08$.    As expected, the overshoot stress increases with
increasing strain amplitude or average strain rate.  Typical strain
rates at $t=0$ are $\dot{\gamma}\tau=7.54 \times 10^{-5}$ for
$\gamma_{0} = 0.06$ and $\dot{\gamma}\tau=2.51 \times 10^{-4}$ for
$\gamma_{0} = 0.20$.   The inset in Fig.\,\ref{fig:stress_strain_T4}
shows the averaged value of the overshoot stress as a function of
the strain amplitude.

\vskip 0.05in


We next perform a more detailed microscopic analysis that involve
spatial configurations of atoms with large relative displacements
during periodic shear.  In general, a combination of a linear
transformation and a translation define the so-called affine
deformation of the material.  In turn, a deviation from the linear
strain field can be described by a nonaffine component of
displacement of atoms relative to their neighbors.  The nonaffine
measure is defined via the transformation matrix $\mathbf{J}_i$ that
best maps all bonds between the $i$-th atom and neighboring atoms
during the time interval $\Delta t$ as follows~\cite{Falk98}:
\begin{equation}
D^2(t, \Delta t)=\frac{1}{N_i}\sum_{j=1}^{N_i}\Big\{
\mathbf{r}_{j}(t+\Delta t)-\mathbf{r}_{i}(t+\Delta t)-\mathbf{J}_i
\big[ \mathbf{r}_{j}(t) - \mathbf{r}_{i}(t)    \big] \Big\}^2,
\label{Eq:D2min}
\end{equation}
where the sum is taken over $N_i$ nearest-neighbor atoms within the
distance $1.5\,\sigma$ from $\mathbf{r}_{i}(t)$. In what follows,
the quantity $D^2(t, \Delta t)$ was evaluated during $\Delta t = T$
with respect to atomic configurations at zero strain.    Note that
in the definition of the nonaffine measure, Eq.\,(\ref{Eq:D2min}),
$D^2$ is normalized by the number of neighbors in the first shell,
and, therefore, the value $D^2 \approx 0.01\,\sigma^2$ corresponds
approximately to a displacement of the atom $\mathbf{r}_{i}$ on the
order of the cage size with respect to its neighbors.

\vskip 0.05in

\begin{figure}[t]
\includegraphics[width=12.cm,angle=0]{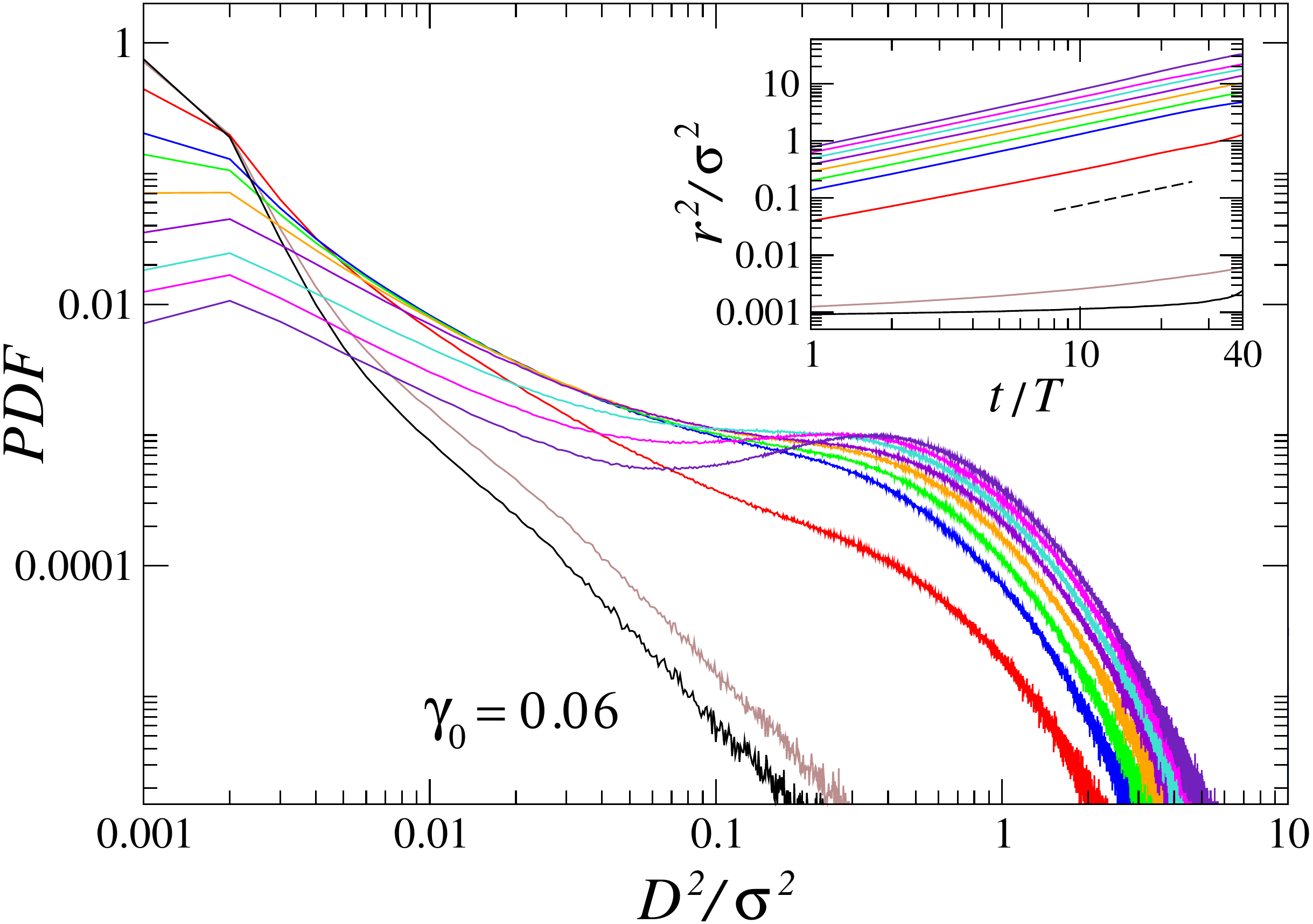}
\caption{(Color online) The normalized probability distribution
function of $D^2(t,T)$ after one cycle for the strain amplitudes
$\gamma_{0} = 0.06$, $0.07$, $0.08$, $0.09$, $0.10$, $0.12$, $0.14$,
$0.16$, $0.18$, and $0.20$ (from left to right). The inset shows the
mean square displacement of atoms for the same strain amplitudes
(from bottom to top). The straight dashed line with unit slope is
plotted for reference. }
\label{fig:d2min_pdf}
\end{figure}

The probability distribution function of the nonaffine measure is
plotted in Fig.\,\ref{fig:d2min_pdf} for the selected strain
amplitudes. The quantity $D^2(t, T)$ was computed after each cycle
with respect to zero strain, i.e., at $t=0, T, ...,\,39\,T$.   The
data in Fig.\,\ref{fig:d2min_pdf} were averaged over five
independent samples and $40$ oscillation cycles.    It is evident
that at the strain amplitudes $\gamma_{0} = 0.06$ and $0.07$, the
distribution function decays rapidly at small values of $D^2$, which
implies that most of the atoms return to their cages after each
cycle. We comment that some atoms with $D^2>0.01\,\sigma^2$ after
one cycle might return back to their cages after several cycles,
which does not necessarily lead to structural relaxation of the
material.  Examples of reversible cage jumps during $50$ oscillation
cycles were reported in the previous MD study~\cite{Priezjev16}.
With increasing strain amplitude, $\gamma_{0} \geqslant 0.08$, the
shape of the probability distribution function becomes more broad,
and it acquires a local maximum at $D^2>0.1\,\sigma^2$ and
$\gamma_{0} \geqslant 0.18$.   The appearance of large nonaffine
displacements at $\gamma_{0} \geqslant 0.08$ indicates a large
number of cage breaking events and it suggests that at least a part
of the material becomes fluidized.   These conclusions are
consistent with the abrupt change of the mean square displacement
from a subdiffusive plateau at small strain amplitudes $\gamma_{0}
\leqslant 0.07$ to a diffusive behavior at $\gamma_{0} \geqslant
0.08$, which is shown in the inset of Fig.\,\ref{fig:d2min_pdf}.

\vskip 0.05in

\begin{figure}[t]
\includegraphics[width=12.cm,angle=0]{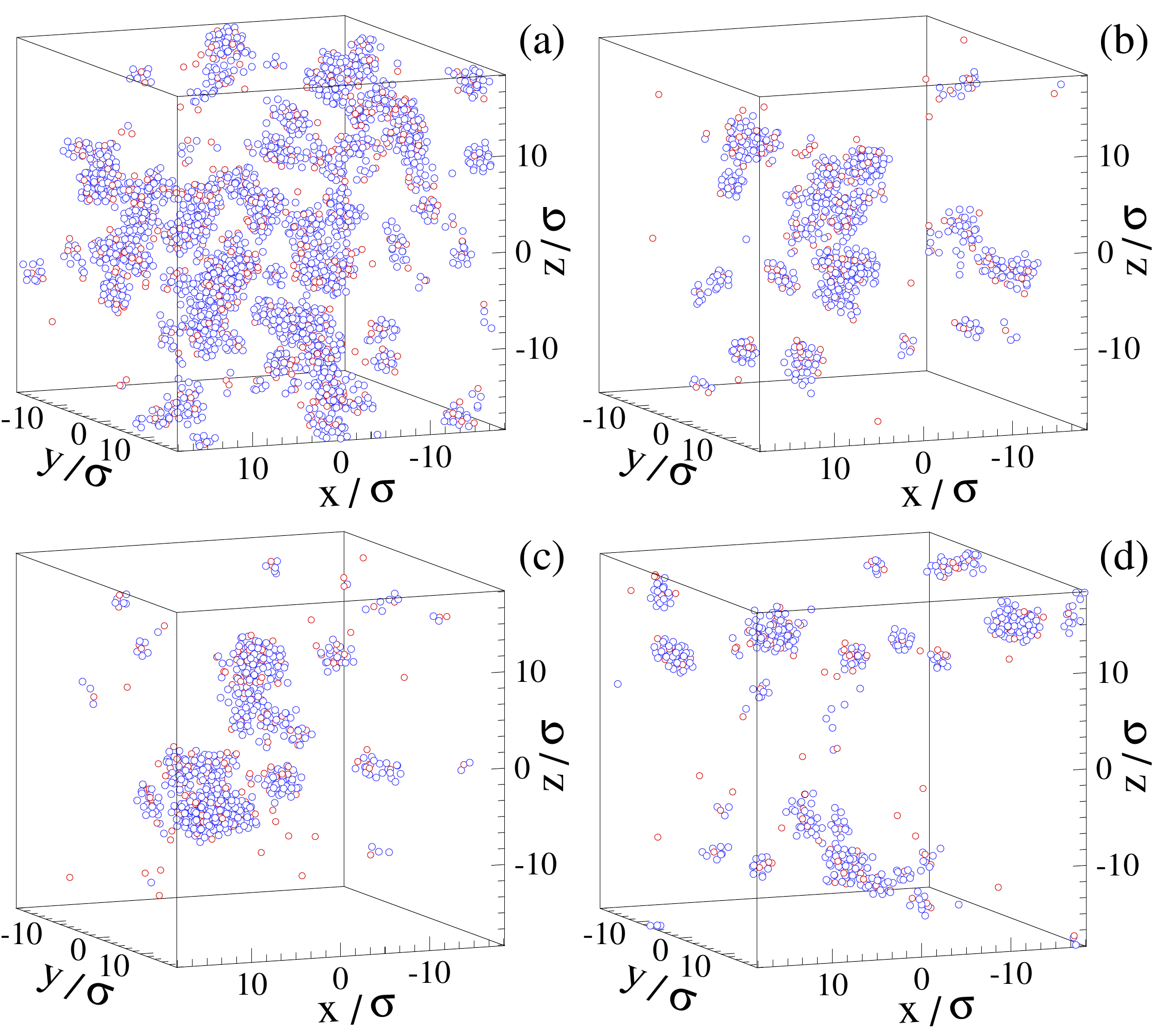}
\caption{(Color online) Instantaneous atomic configurations for the
strain amplitude $\gamma_{0}=0.07$ and the nonaffine measure (a)
$D^2(0,T)>0.01\,\sigma^2$, (b) $D^2(9T,T)>0.01\,\sigma^2$, (c)
$D^2(19T,T)>0.01\,\sigma^2$ and (d) $D^2(39T,T)>0.01\,\sigma^2$. The
atoms of types A and B are marked by blue and red circles,
respectively. }
\label{fig:snapshot_gam_0.07}
\end{figure}
\begin{figure}[t]
\includegraphics[width=12.cm,angle=0]{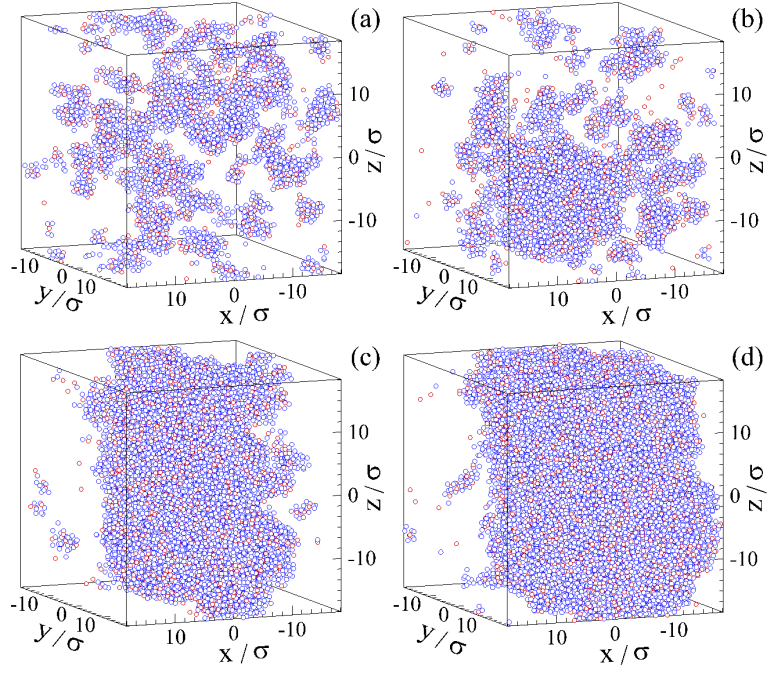}
\caption{(Color online) Snapshots of atom positions for the strain
amplitude $\gamma_{0}=0.08$ and the nonaffine measure (a)
$D^2(0,T)>0.01\,\sigma^2$, (b) $D^2(9T,T)>0.01\,\sigma^2$, (c)
$D^2(19T,T)>0.01\,\sigma^2$ and (d) $D^2(39T,T)>0.01\,\sigma^2$. }
\label{fig:snapshot_gam_0.08}
\end{figure}
\begin{figure}[t]
\includegraphics[width=12.cm,angle=0]{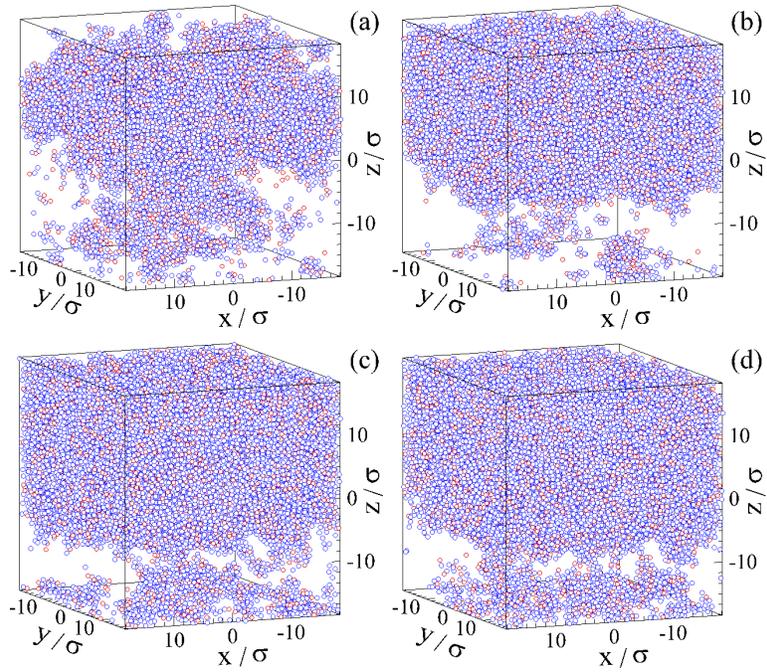}
\caption{(Color online) Configurations of atoms of types A and B for
the strain amplitude $\gamma_{0}=0.10$ and the nonaffine measure (a)
$D^2(0,T)>0.01\,\sigma^2$, (b) $D^2(9T,T)>0.01\,\sigma^2$, (c)
$D^2(19T,T)>0.01\,\sigma^2$ and (d) $D^2(39T,T)>0.01\,\sigma^2$. }
\label{fig:snapshot_gam_0.10}
\end{figure}
\begin{figure}[t]
\includegraphics[width=12.cm,angle=0]{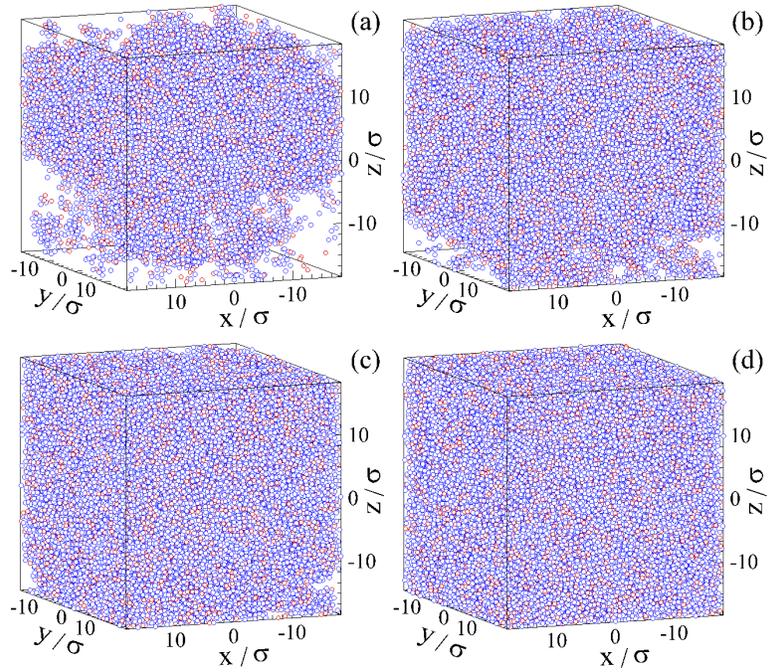}
\caption{(Color online) Positions of atoms A and B for the strain
amplitude $\gamma_{0}=0.16$ and the nonaffine measure (a)
$D^2(0,T)>0.01\,\sigma^2$, (b) $D^2(9T,T)>0.01\,\sigma^2$, (c)
$D^2(19T,T)>0.01\,\sigma^2$ and (d) $D^2(29T,T)>0.01\,\sigma^2$. }
\label{fig:snapshot_gam_0.16}
\end{figure}

Further insight into the structural relaxation process can be gained
by examining the spatial configurations of atoms with large
nonaffine displacements.    Examples of particle positions with
$D^2(t,T)>0.01\,\sigma^2$ are presented in
Figures\,\ref{fig:snapshot_gam_0.07}, \ref{fig:snapshot_gam_0.08},
\ref{fig:snapshot_gam_0.10}, and \ref{fig:snapshot_gam_0.16} for
different strain amplitudes.   In each case, the quantity $D^2(t,T)$
was evaluated for each atom after a full back-and-forth cycle with
respect to the selected reference times that correspond to zero
strain.   It can be observed in Fig.\,\ref{fig:snapshot_gam_0.07}
that during cyclic loading at the strain amplitude $\gamma_{0} =
0.07$, atoms with large nonaffine displacements tend to aggregate
into disconnected clusters.    Interestingly, the number of atoms
with $D^2(0,T)>0.01\,\sigma^2$ is larger after the first cycle,
indicating significant rearrangements of atoms with respect to their
equilibrium positions in the annealed sample, which is followed by a
steady process with smaller scattered clusters.   A similar trend
with the initial decrease of the number of rearrangements with large
irreversible displacements was observed in periodically sheared
suspensions below the strain threshold~\cite{Chaikin08}.

\vskip 0.05in


The most striking observation from the sequence of snapshots shown
in Fig.\,\ref{fig:snapshot_gam_0.08} for the strain amplitude
$\gamma_{0} = 0.08$ is the transition from initial non-percolating
clusters of atoms with $D^2>0.01\,\sigma^2$ to the formation of the
shear band across the whole system after 20th cycle. Note that this
dynamic transition correlates well with the drop in amplitude of the
shear stress oscillations for $\gamma_{0} = 0.08$ shown in
Fig.\,\ref{fig:stress_strain_40T} and with the increase in the
potential energy in Fig.\,\ref{fig:poten}.    Thus, once the
system-spanning shear band is formed, the maximum shear stress
during periodic loading is determined by the fluidized region.   We
observed that the orientation of a shear band is parallel to either
$yz$ or $xy$ planes in different samples.   This is consistent with
the formation of shear bands parallel or perpendicular to the shear
flow direction in binary LJ glasses subjected to a constant strain
rate~\cite{Horbach16}.

\vskip 0.05in


As is evident from Fig.\,\ref{fig:snapshot_gam_0.10}, the shear band
is formed parallel to the $xy$ plane during the first oscillation
cycle at the strain amplitude $\gamma_{0} = 0.10$.   Over the next
40 cycles, the width of the shear band gradually increases, however,
it remains smaller than the system size.    Note that a single shear
band is extended in the $\hat{z}$ direction via periodic boundary
conditions in Fig.\,\ref{fig:snapshot_gam_0.10}\,(c) and (d).
Furthermore, as shown in Fig.\,\ref{fig:snapshot_gam_0.16}, the
shear band appears at the first cycle for the strain amplitude
$\gamma_{0} = 0.16$, and the whole sample becomes fluidized after
30th cycle [see Fig.\,\ref{fig:snapshot_gam_0.16}\,(d)]. With
further increasing strain amplitude, the number of cycles required
for atoms with large nonaffine displacements to be distributed
uniformly throughout the system decreases (not shown).


\vskip 0.05in

Taken together, the results in this study for the potential energy,
stress-strain hysteresis, shear stress, and spatial distribution of
nonaffine displacements indicate that the yielding transition occurs
at strain amplitudes $0.07 < \gamma_{0} < 0.08$.    This is
consistent with the crossover from exponential to power-law decay
(when the strain amplitude is varied from $\gamma_{0} = 0.07$ to
$0.08$) of the spatial correlation function of nonaffine
displacements reported in the previous study~\cite{Priezjev16a}. At
larger strain amplitudes, the cyclic deformation induces
irreversible nonaffine rearrangements of atoms within a shear band
that reduce the amplitude of shear stress oscillations.

\section{Conclusions}

In summary, the dynamic response of binary LJ glasses to oscillatory
shear was investigated using molecular dynamics simulations.    The
periodic shear strain was applied to quiescent samples that were
prepared by cooling down with a computationally slow rate to a
temperature well below the glass transition.   It should be
emphasized that temperature was regulated by the dissipative
particle dynamics thermostat that avoids profile biasing and thus
represents a good choice for studying problems that involve flow
localization.

\vskip 0.05in

It was found that during cyclic loading near the critical strain
amplitude, the relaxation process involves transient,
non-percolating clusters of atoms with large nonaffine displacements
for a number of cycles, which are followed by the formation of a
shear band running across the sample. The appearance of the shear
band causes a noticeable drop in the shear stress amplitude and
marks the onset of irreversible particle rearrangements leading to a
sharp increase of the potential energy.  With increasing strain
amplitude, the stress-strain hysteresis becomes more pronounced and
the plastic flow develops in the whole sample.

\vskip 0.05in

In the future, it will be interesting to determine more precisely
the critical strain amplitude that sets irreversible nonaffine
displacements and to explore its dependence on temperature,
oscillation frequency, system size and preparation history.

\section*{Acknowledgments}

Financial support from the National Science Foundation (CNS-1531923)
is gratefully acknowledged.  The molecular dynamics simulations were
performed using the open-source LAMMPS numerical code~\cite{Lammps}.
Computational work in support of this research was performed at
Michigan State University's High Performance Computing Facility and
the Ohio Supercomputer Center.


%

%

%

%

%

%

%

%

%

%

\bibliographystyle{prsty}

\end{document}